\documentclass[referee]{aa} 
\usepackage{graphicx}
\usepackage{aalongtable}        
\usepackage{natbib}             
\bibpunct{(}{)}{;}{a}{}{,}      
\usepackage{txfonts}
\begin{document}
   \title{Spectroscopic metallicities of Vega-like stars}

   \subtitle{}

   \author{C. Saffe\inst{1}\fnmsep\thanks{On a fellowship from CONICET, Argentina.}
          \and
          M. G\'omez\inst{2}
          \and
          O. Pintado \inst{3}
          \and
          E. Gonz\'alez \inst{4}
          }

   \institute{Complejo Astron\'omico El Leoncito, CC 467,
              5400 San Juan, Argentina\\
              \email{csaffe@casleo.gov.ar}
         \and Observatorio Astron\'omico de C\'ordoba, Laprida 854, 5000
              C\'ordoba, Argentina\\
              \email{mercedes@oac.uncor.edu}
         \and Instituto Superior de Correlaci\'on Geol\'ogica (INSUGEO), 4000 Tucum\'an, Argentina \\
              \email{opintado@tucbbs.com.ar}
         \and Facultad de Ciencias Exactas, F\'\i sicas y Naturales (UNSJ), 5400 San Juan, Argentina \\
              \email{erip.p.a.gonzalez@gmail.com}
             }

   \date{Received September 15, 1996; accepted March 16, 1997}

 
  \abstract
   {}
   {To determine the metallicities of 113 Southern Hemisphere Vega-like candidate stars in relation
   to the Exoplanet host group and field stars.
   }
   {We applied two spectroscopic methods of abundance determinations: equivalent width measurements
together with the ATLAS9 (Kurucz 1993) model atmospheres and the WIDTH9 program, and a comparison of
observed spectra with the grid of synthetic spectra of Munari et al. (2005).
   }
   {For the Vega-like group, the metallicities are indistinguishable from those of field stars
not known to be associated with planets or disks. This result is quite different
from the metallicities of Exoplanet host stars which are metal-rich in comparison to field
stars.
    
   }
   {}

   \keywords{Techniques: spectroscopic -- Stars: abundances -- Stars: late-type}

   \maketitle
%

\section{Introduction}

It is well established that Exoplanet host stars are, on average, metal-rich in
comparison to stars that do not harbor Doppler detected planets 
\citep[see, for example, ][]{santos04}. Two hypotheses have been put forward to explain this
peculiarity of the Exoplanet host stars a) a primordial origin and b) a pollution 
of the convective zone of the star. In the first case the ''excess'' of metallicity
was already present in the parent cloud from which the star bearing planet/s was
formed \citep[see, for example, ][]{santos01}. In the pollution scenario the convective
zone of the star is contaminated by the infall or accretion of planets and/or
planetesimals \citep[see, for example, ][]{gonzalez01}. 

\citet{santos04} found a lack of correlation between the thickness of the
convective zone and the metallicity for a sample of FG dwarfs with planets. 
As the convective zone acts as a diluting medium, for a given amount of accreted
material, F dwarfs with thinner convective zones should exhibit a greater degree
of pollution than G dwarfs with thicker zones. On average, F and G dwarfs exhibit
similar metallicities and the pollution hypothesis is not favored by these
observations. The primordial origin of the ''excess'' remains an alternative to 
explain the relatively high metallicity of stars with planets with respect to
field stars. 

However, \citet{pasquini07} compared the metallicities
of giant and dwarf stars with planets and found that the first group
has, on average, lower metallicities than the dwarfs. The smaller mass of the
convective zone of the dwarfs with respect to the giants provides
a plausible explanation for this difference. The diluting effect of 
the convective zone is efficient for the giants and tends to lower the
metallicity to its primordial value. In this case, the pollution
scenario is favored (over the primordial origin) since it can explain
the observed difference in metallicities between
dwarfs and giants with planets.  Even when the origin or the cause of the ''excess'' of
metallicity of stars with planets is not well understood, the Exoplanet host stars
are metal-rich and this is a feature that distinguishes this group among
stars with similar physical properties and no giant planets detected. 

Vega-like stars are a group of objects that show infrared excesses in their
spectral energy distributions that can be attributed to the presence of dust
in circumstellar disks. The first members or candidate members of the class
were selected by IRAS and had mainly A-F spectral types 
\citep{aumann84,gillett86,backman-paresce93,sylvester96,mannings-barlow98,fajardo-acosta99,
sylvester-mannings00,habing01,laureijs02,sheret04}.
Vega ($\alpha$ Lyr) is one of the four proto-types of the group or the ''fabulous four''
\citep[Vega, $\beta$ Pictoris, Fomalhaut $= \alpha$ PsA and $\epsilon$ Eridanis;][]{gillett86}
and has given the name to the class. 

More recently, Spitzer has contributed
with the detection of G dwarfs with infrared excesses 
\citep{meyer04,rieke05,kim05,chen05,uzpen05,beichman05,beichman06,bryden06,silverstone06,su06,trilling08}.
Since the excesses come from
distances similar to the Kuiper-Belt to the Sun, these stars have also received the designation
of Kuiper-Belt analogs or Kuiper-Belt-like stars. In this contribution we adopt the term
''Vega-like stars'' to refer to both IRAS and Spitzer detections. 

The metallicity of
Vega-like stars has previously been investigated by \citet{greaves06} 
and \citet{chavero06}, deriving nearly solar values. However these works
analyzed relatively small samples of objects. \citet{greaves06}
studied a group of 18 FGK Vega-like stars whereas \citet{chavero06}
included 42 FG dwarfs with infrared excesses in their metallicity determination.
In addition these previous works do not include stars of A spectral type which represent
the bulge of IRAS detections. \citet{greaves06}
derived their sample from the Doppler searches for
planets that in general include solar type stars. \citet{chavero06}
used the Str\"omgren photometry to determine the metallicity. These authors
were also restricted to late spectral types. 

Both the stars with planets and the Vega-like stars have evidence of
the presence of circumstellar material, in the form of planet/s, in the
first case, or dust in a circumstellar disk, in the second. As mentioned
before, the Exoplanet hosts are metal-rich. This fact may have facilitated
the formation of planets \citep{pollack96}. In this contribution 
we determine spectroscopic metallicities of a large sample of Vega-like stars
to compare with the Exoplanet host group. We include objects of B--K 
spectral types, observable from the Southern Hemisphere.


\section{The sample}

We compiled a total of 113 Southern Hemisphere Vega-like candidate stars from the literature,
based on their infrared or submillimetric excess emissions
\citep{backman-paresce93,sylvester96,mannings-barlow98,fajardo-acosta99,sylvester-mannings00,
habing01,laureijs02,sheret04}.
This compilation also includes G dwarfs with infrared excess recently detected by Spitzer
\citep{beichman05,beichman06,bryden06,su06,trilling08}.
Specifically the list comprises objects with BAFGK spectral types
(22, 38, 28, 17 and 8, respectively).  All the stars are luminosity class
V (Hipparcos catalogue) and have distances between 5 and 300 pc.
Table \ref{muestra} lists the observed objects. 

Table \ref{muestra} includes a sub-sample of stars that were originally
selected by IRAS as candidate Vega-like stars. However, when observed
by Spitzer the infrared excesses were deemed to be of little significance.
These objects are: HD 10800, HD 20794, HD 38393, HD 41700, HD 68456, HD 160691,
HD 169830, HD 203608, and HD 216437 
\citep{beichman05,beichman06,bryden06,hill08,trilling08}.
For example, \citet{bryden06} found that for
HD 10800 ${f_{\rm MIPS\_70 \mu m} \over f_*} = 1.3$ 
(the observed flux over the photospheric emission at 70 $\mu$m) and
${f_{\rm MIPS\_70 \mu m} \over f_*} = 1.2$ for HD 68456. This group of objects
should be considered with caution. 


\section{Observations and data reduction}

The stellar spectra were obtained at the Complejo Astronomico El Leoncito (CASLEO), 
using the \emph{Jorge Sahade} 2.15-m telescope
equipped with a REOSC echelle spectrograph\footnote{On loan from the Institute d'Astrophysique
de Liege, Belgium.} and a TEK 1024$\times$1024 CCD detector.
The REOSC spectrograph uses gratings as
cross dispensers. We used a grating with 400 lines mm$^{-1}$, covering
the spectral range $\lambda\lambda$3500--6500, giving a 
resolving power of $\sim$ 12500.  Three individual spectra for each object
were obtained in four observing runs: August 05--08 2005, August 18--22 2005, 
February 18--25 2006 and May 04--07 2007 and have S/N ratio of about 300.

The spectra were reduced using IRAF
\footnote{IRAF is distributed by the National Optical Astronomical Observatories which is operated
by the Association of Universities for Research in Astronomy, Inc., under a cooperative agreement
with the National Science Foundation.} standard procedures for echelle spectra. 
We applied bias and flat corrections and then 
normalized order by order with the {\it continuum} task, using 7--9 order Chebyshev polynomials.
We also corrected by the scattered light in the spectrograph ({\it apscatter} task).
We fitted the background with a linear function on both sides of the echelle apertures,
using the task {\it apall}.  The resolution of the reduced spectra is 0.17 \AA/pix.


\section{Metallicity determinations}

We used two different methods of abundance determination:
1) Fe lines equivalent width measurements together with 
the ATLAS9 \citep{kurucz93} model atmosphere corresponding to a
given star and the WIDTH9\footnote{http://kurucz.harvard.edu/programs.html} program.
2) A comparison of the observed and synthetic spectra using the Downhill method
\citep{gray01}. In particular we used the grid of synthetic spectra calculated by 
\citet{munari05}. This method offers the advantage that there is no need
to identify and measure the equivalent widths of many Fe lines as with the
WIDTH9 program. 

\subsection{Metallicity determinations using the WIDTH program}

To determine abundances by this method it is necessary to
estimate the stellar parameters T$_{\rm eff}$ and Log g, 
by means of the Str\"omgren photometry, for example. With
these quantities we adopt the \citet{kurucz93}'s model atmosphere
appropriated to each star. The model that initially is chosen has 
solar metallicity. Finally the Kurucz's model together with 
the measured equivalent widths are used by the 
WIDTH9 program \citep{kurucz92,kurucz93} to derive the metallicity.

To obtain T$_{\rm eff}$ and Log g, we
have used the uvby$\beta$ mean colors of \citet{hauck98}
with two different calibrations: \citet{napiwotzki93} and \citet{castelli97}
and \citet{castelli98} (hereafter N93 and C97, respectively), with the TEMPLOGG code \citep{rogers95}.
This program has been used in the COROT mission preparation
\citep[see, for example, ][]{lastennet01,gillon-magain06}
and includes reddening corrections, according to \citet{domingo-figueras99}
for stars in the range A3--F0, and to \citet{nissen88}
for spectral types F0--G2. 

We have compared the temperatures and gravities derived using both
calibrations (N93 and C97) and noticed some differences particularly
in the later parameter. For this reason we initially determined metallicities
using values derived from both calibrations and later on considered if they
significantly affect the final metallicity values. We have also confronted the
obtained T$_{\rm eff}$ with those published by \citet{nordstrom04}. We
found a good agreement in particular with the N93 calibration. 
With the values of T$_{\rm eff}$ and Log g
derived for each object, we have chosen the corresponding model atmosphere
using the Kurucz ATLAS9 \citep{kurucz93} code.

The stellar lines were identified using the general references
of {\it A multiplet table of astrophysical interest}
(Moore 1945) and {\it Wavelengths and transition probabilities for atoms and
atomic ions - Part 1: Wavelengths} \citep{reader80},
as well as more
specialized references for the \ion{Fe}{ii} lines \citep{johansson78}.
The equivalent widths were measured by fitting Gaussian profiles through
the stellar metallic lines using the IRAS \emph{splot} task.
There is no more than a 15\% difference among the equivalent widths
of the same lines, measured in different spectra. We have
excluded from our abundance determinations seriously blended lines.

To determine the abundances we need an initial estimation of the microturbulent
velocity ($\xi$). For this estimation we have used the standard method.  We
computed the abundances from the Fe lines for a range of possible values of $\xi$
satisfying two conditions: a) that the abundances of Fe lines were not dependent
on the equivalent widths and b) that the rms errors were minima. To achieve the
first condition the slope in the plot abundance vs $\xi$ must be zero. We tried 
different $\xi$ values to fulfill this requirement. In this sense the
abundance and microturbulent velocity determinations are recursive and 
simultaneous. Once a $\xi$ value has been fixed the corresponding
abundances to all chemical species measured are determined using the WIDTH9 code. 

The WIDTH9 code requires the model atmosphere calculated by the ATLAS9 program, the
equivalent width of each line as well as atomic constants such as 
oscillator strength (Log gf) values, excitation potentials, damping constants, etc. 
In particular for the Log gf we used \citet{fuhr88} and \citet{kurucz92}. 
This code calculates the theoretical equivalent widths for an initial 
input abundance and compares these values with the measured equivalent widths. 
Then the code modifies the abundance to achieve a difference between theoretical
and measured equivalent widths $<$ 0.01 m\AA.
The final values of the metallicities corresponding to the N93 and C97 calibrations,
are listed in Table \ref{width.metal}. We have included the number of lines used in each 
determination as well as the rms of the average. 

To estimate errors for our WIDTH metallicities we consider the following 
facts. The most significant contribution to the final uncertainties, probably,
comes from the equivalent width measurements. We assume a 5\% error due to 
the continuum level determination. This translates into 20\% maximum uncertainties
in the metallicity estimation. The atomic constants may also have
uncertainties. In particular we estimate that the oscillator strength
values may cause differences of about 10\% in the calculated metallicity.
Finally to provide an estimation of ''typical'' errors introduced by
the WIDTH method we increased the T$_{\rm eff}$ by 150 K and the Log gf by 0.15, 
and recalculated the metallicity value for each star. We derived a median
difference of 0.20 dex. The largest difference corresponds to HD 28978 
(0.55 dex). 

\subsection{Metallicity derivations from synthetic spectra: The Downhill method}

The WIDTH method is not practical when the number of stars is large. 
For each object, we need to identify and measure many spectral lines. An alternative
would be to compare the observed spectra with a grid of synthetic ones corresponding
to different values of the metallicites and choose from the grid the spectrum that better
reproduces the observed data \citep{gray01}. This comparison has the advantage
that the complete profiles of the lines and not only the equivalent widths are used
in the metallicity determinations. 

In general synthetic spectra depend on four parameters: T$_{\rm eff}$, surface gravity (Log g), 
metallicity ([Fe/H]) and microturbulent velocity ($\xi$).  Following \citet{gray01}, we 
applied a multidimensional Downhill Simplex
technique, in which the observed spectrum is compared to a grid of synthetic spectra.
The ''final'' synthetic spectrum is an interpolation of spectra, rather than
a single point in the grid. As we are working with four variables (T$_{\rm eff}$, Log g, [Fe/H] and
$\xi$) the interpolation is done in 4d, minimizing the square differences in each wavelength
(i.e., the $\chi^2$ statistics). The stellar parameters are
determined with a higher accuracy than the steps in the grid since
they correspond to interpolated values. 

The grid of synthetic spectra was taken from \citet{munari05}. 
The parameters range covered by the grid is the following: 

\[
\begin{array}{lp{0.8\linewidth}}
& 3500 K $<$ T$_{\rm eff}$ $<$ 40000 K, with steps of 250 K,   \\
& 0.0 dex $<$ Log g $<$ 5.0 dex, with steps of 0.5 dex,        \\
& $-$2.5 dex $<$ [Fe/H] $<$ 0.5 dex, with steps of 0.5 dex,    \\
& and $\xi$ values of 0, 1, 2, and 4 km/s.  \\
\end{array}
\]

\noindent
In addition to these parameters, the synthetic spectra are
calculated for 15 different rotation velocities,
ranging 0 -- 500 km/s. In all, \citet{munari05}'s 
library contains 625000 different spectra. These authors
calculated the complete synthetic spectral library for
four resolving powers: 20000, 11500 (GAIA), 8500 (RAVE) and
2000 (SLOAN). To our request, Dr. U. Munari kindly provided
a grid corresponding to the REOSC/CASLEO resolving power (12500). 

Synthetic spectral lines were convolved with the instrumental line
profile corresponding to the REOSC/CASLEO. Finally they were also
convolved with a Gaussian profile corresponding to the rotational
velocities of the sample stars, taken from the literature
\citep{glebocki00,mora01,yudin01,royer02,cutispoto02,cutispoto03,pizzolato03,strom05,reiners06}.
We weighted the synthetic spectra by the blaze function
of each of the REOSC spectrograph order. Finally we normalized  
and re-sampled our data to compare them with \citet{munari05}'s grid.
The spectral sampling of the synthetic spectra is 0.02 \AA. 

We have implemented the Downhill method \citep{gray01} by means
of a Fortran program. From the stellar spectral type or the 
Str\"omgren photometry it is possible to estimate ''a starting
point'' in the 4d grid. The Downhill method provides a searching
algorithm within the 4d grid and finds the best match, minimizing
the $\chi^2$. In our case, the final spectrum is obtained by an interpolation
of 16 spectra of \citet{munari05}'s grid. In general it takes
15 -- 20 min for each star (50 -- 60 iterations) in a Pentium IV 2.0 GHz
to find the best interpolated spectrum. Table \ref{downhill.metal} lists the metallicities
obtained with the Downhill method for our sample of Vega-like stars.

To estimate the uncertainties in the metallicities obtained by the
Downhill method, we carried out a few tests. We first applied this
method to 30 synthetic spectra of known metallicities. The median
difference between the derived and known metallicities is 0.2 dex. 

The internal consistency of the method has been checked,
by fixing one of the four variables and comparing the
resultant metallicities. Fixed values for each 
variable were obtained, for example, from an adopted 
calibration: 

\[
\begin{array}{lp{0.8\linewidth}}
& a) T$_{\rm eff}$ was taken from the N93 calibration, \\
& b) Log g was adopted from the N93 calibration, \\
& c) $\xi$ was fixed at 2.9 km/s, the solar value. \\
\end{array}
\]

\noindent
The median difference, calculated by fixing 3 of the 4 variables 
with respect to the ''standard'' procedure (i.e., with 4
variable), was 0.05 dex. Considering this value and the 
median difference derived from the comparison with 30
synthetic spectra of known metallicities (0.2 dex), we 
estimate a ''typical'' uncertainty of 0.06 dex for the
metallicities derived by the Downhill method. 

We have also compared the Downhill method derived metallicities
with those obtained by \citet{nordstrom04} and
\citet{fischer-valenti05}. We first noticed a systematic
difference of $\sim$ 0.09 dex between these two determinations. 
\citet{fischer-valenti05}'s determinations are, on average, 
larger than those from \citet{nordstrom04}'s. Our
Downhill method derived metallicities show a better agreement
with \citet{nordstrom04}'s value than with \citet{fischer-valenti05}'s. 
However this later comparison is based on a relatively small
number of common stars.  

In the work of \citet{nordstrom04}'s the metallicities are
derived as a secondary parameter obtained photometrically. In 
the case of \citet{fischer-valenti05}, the metallicities are obtained
by a comparison with synthetic spectra but using only a small
range of wavelengths (6000 -- 6200 \AA). With these
limitations in mind, we consider that the external consistency of the
Downhill method derived metallicities is acceptable. 

We finally mention two parameters taken as fixed by the
Downhill method, the radial and the rotational velocities. 
Radial velocities are initially determined, 
minimizing the $\chi^2$ with an accuracy of 0.1 km/s or
a median value of 0.03 dex in metallicity. Rotational 
velocities ({\it{v sin i}}) from the literature have ''typical''
dispersions of 5 -- 10\%, corresponding to an error of about
10\% in metallicity.

In summary, we have estimated an internal uncertainty of 0.06 dex for
metallicities derived from the Downhill method. A more conservative
estimation would indicate a value of 0.1 dex. This corresponds to
half of the uncertainty calculated for the WIDTH method (0.2 dex). 
In this manner, the Downhill method allows a more precise determination
of the metallicities for our sample of Vega-like objects. 

\subsection{Comparison of metallicity determinations by the WIDTH and the Downhill methods}

Table \ref{medianas.width.Downhill} lists the medians and the dispersions of the 
metallicities derived by applying the WIDTH and the Downhill methods for the Vega-like
group. In the case of the WIDTH method we present the results corresponding to 
the two calibrations used (N93 and C97). The derived median values are practically 
indistinguishable.

\setcounter{table}{3}
\begin{table}
\center
\caption{Medians and dispersions of the metallicities for the Vega-like sample}
\vskip 0.2in
\begin{tabular}{lccc}
\hline
\hline
Method    & Median & Dispersion & N    \\
          & [Fe/H] & [Fe/H]     &      \\
\hline
WIDTH$+$N93    & $-$0.14 & 0.28 & 113  \\
WIDTH$+$C97    & $-$0.11 & 0.26 & 113  \\
Downhill       & $-$0.11 & 0.27 & 113  \\
\hline
\end{tabular}
\vskip 0.2in

Note - N93: \citet{napiwotzki93}'s calibration; C97: \citet{castelli97}
and \citet{castelli98}'s calibration.
\label{medianas.width.Downhill}
\end{table}

\begin{figure}
\centering
\includegraphics[width=9cm]{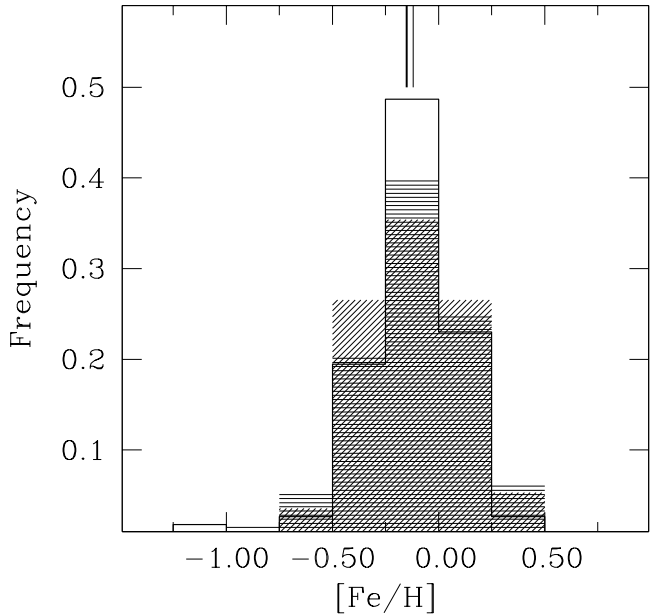}
\caption{Metallicity distributions for the Vega-like sample. Histograms shaded
at 0 and 45 degrees correspond to the WIDTH method derivations using  
N93 \citep{napiwotzki93} and C97 \citep{castelli97,castelli98} calibrations, 
respectively.  The empty histogram shows the metallicity distribution derived by the
Downhill method.  The vertical lines indicate the medians of each distribution.
The left line corresponds to the WIDTH$+$N93 median, and the right line shows (superimposed)
the WIDTH$+$C97 and Downhill medians (see Table \ref{medianas.width.Downhill}).}
\label{3histograms}
\end{figure}

Figure \ref{3histograms} compares the metallicity distributions calculated with
the WIDTH method plus the N93 calibration (histogram shaded at 0 degree) and
the C97 calibrations (histogram shaded at 45 degrees), respectively.  The empty
histogram shows the distribution derived with the Downhill method for
the Vega-like sample. Vertical lines indicate the medians of each 
distribution. The left line corresponds to the WIDTH$+$N93 median, and the right
line shows (superimposed) the WIDTH$+$C97 and Downhill medians
(see Table \ref{medianas.width.Downhill}).
The KS-test \citep{press92} indicates that these distributions are similar
and represent the same parent population. 

We finally adopt the metallicities calculated with the Downhill method for the sample
of Vega-like stars, as these determinations use the complete line profiles and 
not only the equivalent widths. In addition ''typical'' uncertainties are smaller than 
those estimated for the WIDTH method. 

\section{Discussion of the results}

The metallicity of the Solar Neighborhood is usually represented by a
control sample of stars, which should exclude, in our case, known Vega-like stars.
The selection of the control sample is important, because different groups of objects
(i.e., stars selected by different criteria) may have different metallicities.
For example, \citet{fischer-valenti05} compared two different control samples, with the metallicity
distribution of Exoplanet host stars.  Their control sets are volume-limited and
magnitude-limited. The medians of the metallicity ''excess'' of the Exoplanet host stars
compared with the two  groups, are 0.13 and 0.226 dex, respectively. In other words,
the ''excess'' is real, but the amount depends on the control sample used. The
two control sets contain different classes of stars.  The magnitude-limited sample
includes more massive and metal-rich stars than the volume-limited set. 

\begin{figure}
\centering
\includegraphics[width=9cm]{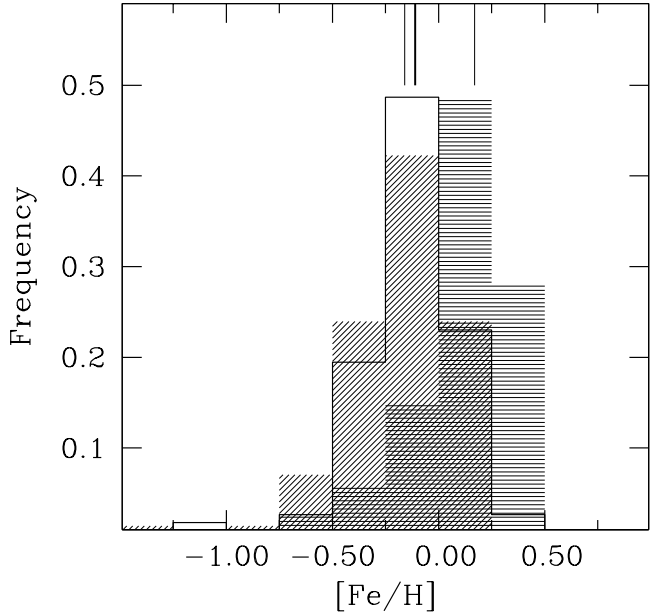}
\caption{Metallicity distributions for the Vega-like sample, empty histogram, 
for stars with planets, histogram shaded at 0 degree, and for stars known
not to harbor planets, histogram shaded at 45 degrees. The vertical lines indicate
the medians of each distribution: stars without planets, Vega-like stars, and
exoplanet host stars, respectively (see Table \ref{medianas2}).}
\label{3comparison}
\end{figure}

The metallicity distribution of Exoplanet host stars is usually compared
with a volume-limited group of solar neighborhood stars 
\citep{gonzalez98,gonzalez99,gonzalez01,santos00,santos03,santos04,sadakane02,laws03}.
We compared the metallicity distribution of 
our Vega-like sample with a volume-limited sample
of 71 stars, without Doppler detected Exoplanets \citep{santos01,gilli06}
and with 98 Exoplanets host stars \citep{santos04}. Metallicity values for these
two comparison samples were obtained from \citet{nordstrom04}. 
As discussed in Section 4.2, the agreement between our metallicities and
those obtained by these authors is acceptable.
Figure \ref{3comparison} shows these distributions. 
Vega-like stars are represented by the empty histogram, stars with planets by 
the histogram shaded at 0 degree and stars known not to harbor planets detected by the 
Doppler technique, by the histogram shaded at 45 degrees. 
The KS test shows no significant difference between the metallicities distributions
of the Vega-like stars and stars without planets. On the other hand, the Vega-like stars
metallicity distribution is different from the metallicity distribution
for stars with planets with a high level of confidence. 

\citet{fischer-valenti05} obtained that the probability
that a FGK star harbors a giant planet/s increases as
P(Z) $\propto$ $(10^{\rm Z})^2$, where Z is the stellar metallicity
\citep[see also, ][]{wyatt07b}.
If this relation is also applicable to
A stars (the bulge of IRAS detected Vega-like stars), the low median
value of the metallicity for the Vega-like group ($-$0.11 dex, see
Table \ref{medianas2}) indicates that the probably for these stars to host a planet/s
of the type detected by radial velocity surveys is also low.
We note, however, that the dispersion of metallicities within the
Vega-like stars is also significant (0.26 dex) and at least a fraction
of these stars has metallicities high enough to host giant planets, 
assuming the ''excess'' of metallicity/presence of a giant planet/s holds
for A spectral type stars. 
In addition it is worthwhile to mention that Doppler searches do not
achieve the required precision to detect planets in A stars as metal
lines practically disappear. 

\begin{figure}
\centering
\includegraphics[width=9cm]{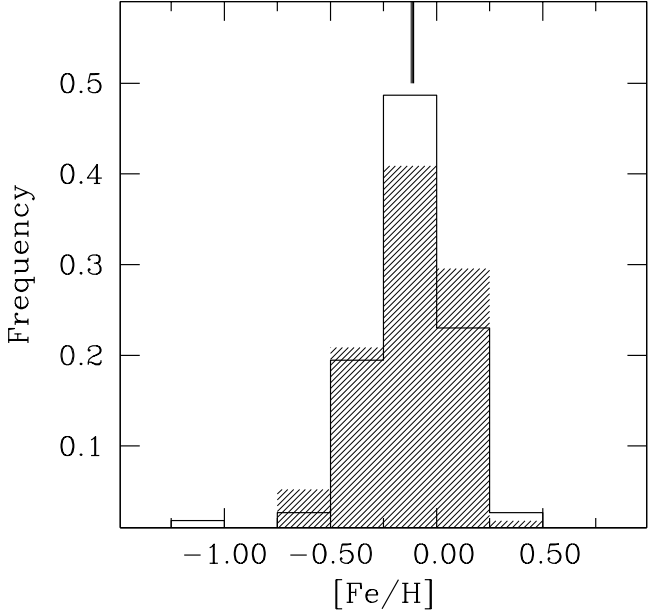}
\caption{Metallicity distributions for the Vega-like sample, empty histogram,
and for stars without excess at 24 or 70 $\mu$m
\citep{beichman05,beichman06,bryden06,su06}.
The vertical lines (almost superimposed) indicate the medians of each distribution.}
\label{2comparison}
\end{figure}

We also compared the metallicity distribution of Vega-like stars, with a sample of 
115 stars without excess at 24 or 70 $\mu$m, observed by Spitzer
\citep{beichman05,beichman06,bryden06,su06}.
Figure \ref{2comparison} shows
these distributions. Vega-like stars are indicated by the empty histogram whereas the stars
without excess at 24 or 70 $\mu$m are shown by the histogram shaded at 45 degrees. 
The KS test shows no significant
difference between the two distributions. 
Table \ref{medianas2} lists the medians and the dispersions of the four
samples compared in Figures \ref{3comparison} \& \ref{2comparison}. 

\begin{table}
\center
\caption{Medians and dispersions of the Vega-like sample and three comparison groups}
\vskip 0.2in
\begin{tabular}{lccc}
\hline
\hline
Sample    & Median & Dispersion & N  \\
          & [Fe/H] & [Fe/H] \\
\hline
Vega-like stars                      & $-$0.11 & 0.27 & 113 \\
Exoplanet host stars                 & $+$0.17 & 0.22 & 98 \\
Volume-limited sample without planets & $-$0.16 & 0.25 & 71 \\
Stars without excess at 24 or 70 $\mu$m    & $-$0.12 & 0.24 & 115 \\
\hline
\end{tabular}
\label{medianas2}
\end{table}

The results in Table \ref{medianas2} indicate that, on average, the Vega-like group
has metallicities similar to the stars in the Solar Neighborhood without detected
planets or disks, in contrast to the Exoplanet host stars group. This
result confirms and extends previous works by \citet{greaves06}
and \citet{chavero06}, based on relatively small numbers of FG Vega-like
stars.

\begin{figure}
\centering
\includegraphics[width=9cm]{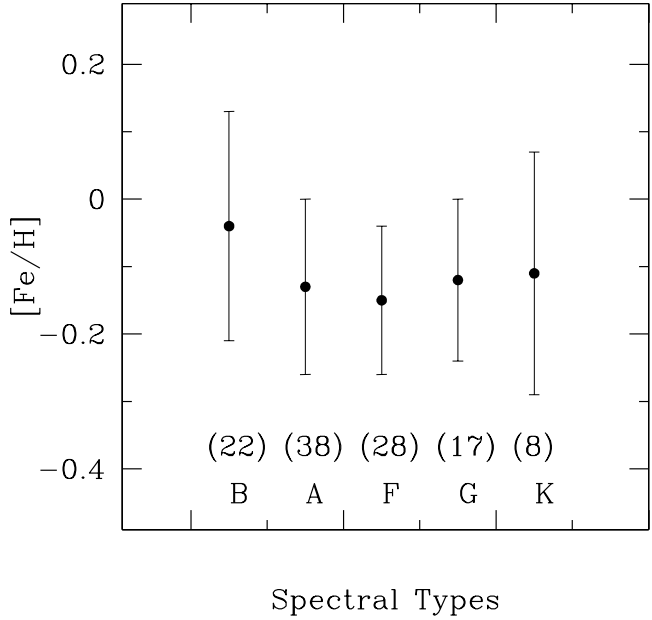}
\caption{Metallicity of Vega-like stars of different spectral types. 
Between brackets is indicated the number of objects in each spectral
type bin. The vertical lines are the corresponding dispersions.}
\label{metalicidad-tipos}
\end{figure}

In Figure \ref{metalicidad-tipos} we analyze the metallicity distribution
of Vega-like stars of different spectral types. The number of
objects corresponding to each spectral type is indicated between 
brackets. The vertical bars are the dispersions within  
the spectral types. A-spectral-type stars still dominate the Vega-like group
although Spitzer has significantly contributed with F and G stars
during the last few years \citep{beichman05,beichman06,bryden06,su06}.
Figure \ref{metalicidad-tipos} shows no trend
of the metallicity with the spectral type for the Vega-like group. 

As suggested by \citet{greaves06} the relatively high metallicity
of Exoplanet host stars as well as the solar metallicity value for the
Vega-like stars can be understood within the core accumulation model of
\citet{pollack96}.
The high metal content of the disk favors the
fast formation of giant planets, which needs to accrete an atmosphere
and migrate inward before the gas is dissipated from the disk. On the
contrary, for Vega-like objects no giant planet needs to be formed and/or
migrate inward. The gas may dissipate and still the planetesimal in the
external part of the disk may produce dust by collisions.

We tentatively analyzed two small sub-sets of Vega-like objects: the Vega-like
stars with planets and the Vega-like group with no Doppler detected planets. 
The first group is composed of 7 stars: 6 with 70 $\mu$m excess detected
by Spitzer 
\citep[HD 33636, HD 50554, HD 52265, HD 82943, HD 128311 and HD 117176; ][]{beichman06}
and $\epsilon$ Eri with infrared and submillimieter
excesses \citep{greaves98,zuckerman01}.
In the second group we include 5 stars without Exoplanets detected by the Doppler
technique \citep{santos04,gilli06} and showing infrared excess in 24
or 70 $\mu$m  \citep[HD 7570, HD 38858, HD 69830, HD 76151 and HD 115617; ][]{beichman06,bryden06}.

The median metallicity of Vega-like stars with planets is $+$0.07 dex and
the dispersion is 0.16 dex. For the Vega-like objects without planets these
values are: $-$0.08 and 0.18 dex, respectively. It seems that when 
a Vega-like star has a planet the metallicity increases slightly. However
the small number of objects available as well as the dispersions prevent
us from giving any statistical significance to this initial trend. 

\citet{greaves07} proposed that the solid-mass (i.e., metals) content
in primordial disks, called M$_{\rm S}$, is the fundamental parameter that
regulates the planet/disk formation.
If M$_{\rm S}$ is small, the star will form a Vega-like disk, while if M$_{\rm S}$
is larger, a giant planet may be formed.  Table 1 of \citet{greaves07}
shows the range of metallicity and the final configurations
(planet$+$debris, debris, etc.) derived by these authors. 
The medians of the metallicities of Vega-like stars with and without planets
agree with \citet{greaves07}'s Table 1. However this can only be considered
as an initial trend that needs to be confirmed by increasing the number of 
Vega-like objects with planets as well as objects known not to harbor Doppler detected
planetary mass objects. 

\section{Summary and Conclusions}

We derived spectroscopic metallicities for a group of 113 Southern Hemisphere
Vega-like stars. We applied two methods to determine metallicities: the ''classical''
WIDTH method and a comparison with the grid of synthetic spectra of \citet{munari05} by means of the
Downhill algorithm. The later method offers the advantage that the complete  
profile of the line is used in the metallicity derivation and not only the equivalent
width. In addition we estimated smaller uncertainties in the metallicities
derived by the Downhill method (0.1 dex) than with the WIDTH code (0.2 dex). 

Vega-like stars have metallicities similar to Solar Neighborhood stars without
planets or disks and significantly different from the Exoplanet host stars.
This result confirms previous estimations by \citet{greaves06} and \citet{chavero06},
based on comparatively smaller samples.

The low metallicities for the Vega-like group (median $=$ $-$11 dex) in
relation to the Exoplanet host stars 
\citep[median $= +$ 0.17, see for example,][]{fischer-valenti05}, may indicate
that the probability for these stars to host a planet/s
of the type detected by radial velocity surveys is also low.
However the dispersion of metallicities within the
Vega-like stars is also significant (0.26 dex) and thus
a fraction of these objects may have metallicities high
enough to form giant planets. We caution that 
Exoplanet host stars are mainly of FGK spectral types 
whereas the bulge of IRAS detected Vega-like stars has
A spectral type which are, in general, excluded from radial
velocity searches since high precisions are not feasible. 
In this we are assuming that the probability of a A star to
be associated with a giant planet depends on the metallicity
as is the case for FGK stars. 

We find no trend in the metallicities of
Vega-like objects with the spectral type.  \citet{greaves06} suggestion
make compatible the relative high metallicity of Exoplanet host stars and the solar 
Neighborhood value for Vega-like stars with the core accumulation model of 
\citet{pollack96}.

Analyzing two relatively small sub-samples, we find that Vega-like
stars with a Doppler detected planet have slightly higher metallicities
than Vega-like stars known not to harbor such a planet. However this
must be considered only as an initial trend that needs to be confirmed by
increasing both samples to achieve a statistical significant result. 

\begin{acknowledgements}
The authors thank Drs. F. Castelli, P. Bonifacio and L. Sbordone
for making their codes available to them.
\end{acknowledgements}

\setcounter{table}{0}
\begin{longtable}{lcccl}
\caption{Sample of Vega-like stars observed at the CASLEO} 
\\
\hline \hline
Star  & Distance & V & Spectral & Reference \\
      & [pc]     &   & Type     & \\
\hline
\endfirsthead
\caption{Continued.}\\
\hline \hline
Star  & Distance & V & Spectral & Reference \\
      & [pc]     &   & Type     & \\
\hline
\endhead
\hline
\endfoot
HD 105     &       40 &     7.51 &      G0V & DEC03 HILL08 \\
HD 142     &       26 &     5.70 &     G1V  & BE05 TR08       \\
HD 2623    &      365 &     7.93 &       K2 & SB91         \\
HD 3003    &       46 &     5.07 &      A0V & MB98 O92 SB91 WY07 \\
HD 9672    &       61 &     5.62 &      A1V & MB98 O92 SB91 PW91 WW88 SN86 WY07 \\
HD 10647   &       17 &     5.52 &      F8V & MB98 O92 SB91 DEC03 TR08 \\
HD 10700   &        4 &     3.49 &      G8V & MB98 HDJL01 DEC03 \\
HD 10800   &       27 &     5.88 &      G2V & MB98 BE06 BR06 \\
HD 17206   &       14 &     4.47 &     F5V  & O92 SB91       \\
HD 17848   &       51 &     5.25 &      A2V & MB98       \\
HD 18978   &       26 &     4.08 &      A4V & SCBS01   \\
HD 20010   &       14 &     3.80 &      F8V & O92 WW88 \\
HD 20794   &        6 &     4.26 &      G8V & DEC03  BE06 \\
HD 21563   &      182 &     6.14 &     A4V  & MB98  \\
HD 22049   &        3 &     3.72 &      K2V & SB91 WW88 HDJL01 DEC03 \\
HD 22484   &       14 &     4.29 &      F9V & DEC03 TR08 \\
HD 23362   &      309 &     7.91 &       K2 & SB91 \\
HD 25457   &       19 &     5.38 &      F5V & DEC03 PAS06 HILL08 \\
HD 28375   &      118 &     5.53 &      B3V & O92 SB91 TR08 \\
HD 28978   &      125 &     5.67 &     A2V  & BP93  \\
HD 30495   &       13 &     5.49 &      G3V & HDJL01 DEC03 TR08 \\
HD 31295   &       37 &     4.64 &      A0V & SN86 WY07 \\
HD 33262   &       12 &     4.71 &      F7V & BR06 TR08 \\
HD 33636   &       29 &     7.00 &       G0 & BE05 TR08  \\
HD 33949   &      172 &     4.36 &      B7V & MB98 O92 SB91 PW91 SN86\\
HD 35850   &       27 &     6.30 &     F7V  & DEC03 PAS06 AP08 \\
HD 36267   &       88 &     4.20 &      B5V & BP93             \\
HD 37484   &       60 &     7.26 &      F3V & PAS06 HILL08    \\
HD 38206   &       69 &     5.73 &      A0V & MB98 DEC03 WY07  \\
HD 38385   &       53 &     6.25 &      F3V & MB98 \\
HD 38393   &        9 &     3.59 &      F7V & MB98 HDJL01 SH03 BE06 \\
HD 38678   &       22 &     3.55 &     A2V  & MB98 O92 PW91 AP91 C87 HDJL01 DEC03 SU06 WY07\\
HD 39014   &       44 &     4.34 &      A7V & O92 SB91 C87 JU04  \\
HD 39060   &       19 &     3.85 &      A3V & MB98 O92 C92 SB91 PW91 AP91 WW88 C87 JJE86 HDJL01 DEC03 WY07 \\
HD 40136   &       15 &     3.71 &      F1V & MB98 BE06b \\
HD 41700   &       27 &     6.35 &     G0V  & DEC03  HILL08 \\
HD 41742   &       27 &     5.93 &      F4V & MB98 \\
HD 43955   &      305 &     5.51 &     B3V  & MB98 \\
HD 66591   &      166 &     4.81 &      B3V & MB98\\
HD 68456   &       21 &     4.74 &      F5V & BR06 TR08 \\
HD 69830   &       13 &     5.95 &      K0V & MB98 BE06 BR06 \\
HD 71043   &       73 &     5.89 &      A0V & WY07 \\
HD 71155   &       38 &     3.91 &      A0V & PW91 C87 WY07 \\
HD 75416   &       97 &     5.46 &     B9V  & MB98 SU06 WY07 \\
HD 76151   &       17 &     6.01 &      G3V & BE06 BR06 \\
HD 79108   &      115 &     6.14 &      A0V & WY07 \\
HD 80950   &       81 &     5.86 &      A0V & MB98 WY07 \\
HD 82943   &       27 &     6.54 &       G0 & BE05 TR08 \\
HD 86087   &       98 &     5.71 &      A0V & BP93 \\
HD 88955   &       32 &     3.85 &      A2V & MB98 \\
HD 98800   &       47 &     8.89 &      K4V & MB98 SB91 WW88 MA05 \\
HD 99211   &       26 &     4.06 &      A9V & MB98\\
HD 102647  &       11 &     2.14 &     A3V  & O92 C92 SB91 PW91 WW88 C87 HDJL01 DEC03 \\
HD 105211  &       20 &     4.14 &      F2  & BE06b  \\
HD 105686  &      101 &     6.16 &      A0V & MB98 \\
HD 108257  &      123 &     4.82 &     B3Vn & BP93  \\
HD 108483  &      136 &     3.91 &      B3V & MB98 \\
HD 109085  &       18 &     4.30 &      F2V & MB98 SB91 SH03 BE06b \\
HD 109573  &       67 &     5.78 &      A0V & TE00 \\
HD 111786  &       60 &     6.14 &      A0  & WY07 \\
HD 113766  &      131 &     7.48 &     F5V  & MB98 O92 CH06 \\
HD 115617  &        9 &     4.74 &      G5V & BR06 \\
HD 115892  &       18 &     2.75 &      A2V & MB98 SU06 WY07 \\
HD 117176  &       18 &     4.97 &      G5V & BE05 BR06 \\
HD 117360  &       35 &     6.52 &      F6V & MB98     \\
HD 121847  &      104 &     5.20 &      B8V & MB98 PW91\\
HD 123160  &          &     8.66 &       K5 & SB91   \\
HD 124771  &      169 &     5.06 &      B4V & MB98  \\
HD 128311  &       17 &     7.48 &       K0 & BE05 BE06        \\
HD 131885  &      121 &     6.91 &      A0V & MB98 \\
HD 135344  &       78 &     7.91 &      F3V & MB98 O92 WW88\\
HD 136246  &      143 &     7.18 &      A1V & WY07 \\
HD 139365  &      136 &     3.66 &    B2.5V & MB98\\
HD 139664  &       18 &     4.64 &     F5V  & O92 PW91 WW88 HDJL01 DEC03 CH06 BE06b \\
HD 141569  &       99 &     7.11 &       B9 & O92 SB91 WW88 JJE86 SH03 CL03 \\
HD 142096  &      109 &     5.04 &      B3V & MB98 O92 SB91 \\
HD 142114  &      133 &     4.59 &   B2.5V  & MB98 O92 \\
HD 142165  &      127 &     5.38 &      B5V & MB98\\
HD 144432  &      253 &     8.19 &     F0V  & MB98 O92 WW88 \\
HD 145482  &      143 &     4.58 &      B2V & MB98 \\
HD 150638  &      240 &     6.46 &      B8V & PW91 \\
HD 152391  &       17 &     6.65 &      G8V & DEC03 BR06 TR08 \\
HD 158643  &      131 &     4.78 &      A0V & O92 \\
HD 158793  &          &     8.83 &          & BP93 \\
HD 159082  &      152 &     6.42 &    B9.5V & BP93 \\
HD 160691  &       15 &     5.12 &      G5V & MB98 BE05 \\
HD 161868  &       29 &     3.75 &      A0V & O92 C87 SN86\\
HD 164249  &       47 &     7.01 &      F5V & PW91 DEC03 \\
HD 164577  &       81 &     4.42 &     A2V  & WA95 \\
HD 165341  &        5 &     4.03 &     K0V  & DEC03 \\
HD 166841  &      214 &     6.32 &      B9V & MB98\\
HD 169830  &       36 &     5.90 &      F8V & BE05 \\
HD 176638  &       56 &     4.74 &     A0V  & MB98\\
HD 177817  &      274 &     6.00 &      B7V & DEC03 \\
HD 178253  &       40 &     4.11 &     A0V  & MB98 PW91 \\
HD 181296  &       48 &     5.03 &     A0V  & BP93 MB98 \\
HD 181327  &       51 &     7.04 &     F6V  & MB98 SCH06 CH06  \\
HD 181869  &       52 &     3.96 &      B8V & MB98 WY07 \\
HD 183324  &       59 &     5.79 &      A0V & WY07 \\
HD 185507  &      209 &     5.18 &     B3V  & BP93 FR96 \\
HD 188228  &       33 &     3.97 &      A0V & SU06 \\
HD 191089  &       54 &     7.18 &      F5V & MB98 HILL08 CH06 \\
HD 198160  &       73 &     5.67 &       A2 & RI05 \\
HD 199260  &       21 &     5.70 &      F7V & BE06b  \\
HD 203608  &        9 &     4.21 &      F6V & MB98 BE06 BR06  \\
HD 206893  &       39 &     6.69 &      F5V & DEC03 \\
HD 207129  &       16 &     5.57 &      G2V & MB98 O92 WW88 O86 HDJL01 DEC03 SH03 TR08 \\
HD 209253  &       30 &     6.63 &     F7V  & DEC03 PAS06 HILL08 \\
HD 216435  &       33 &     6.03 &     G3V  & BP93 \\
HD 216437  &       27 &     6.04 &     G4V  & BE06 BR06 \\
HD 216956  &        8 &     1.17 &      A3V & MB98 O92 C92 SB91 WW88 C87 HDJL01 DEC03 HO98\\
HD 221853  &       71 &     7.35 &       F0 & DEC03  \\
HD 224392  &       49 &     5.00 &      A1V & MB98 O92\\
\label{muestra}
\end{longtable}

\noindent
Note - The distances, visual magnitudes and spectral sypes are taken from the Hipparcos catalog.\\
References (alphabetically sorted):
AP08 = \citet{apai08}, 
BE05 = \citet{beichman05}, 
BE06 = \citet{beichman06},
BE06b = \citet{beichman06b}, 
BP93 = \citet{backman-paresce93}, 
BR06 = \citet{bryden06}, 
C87 = \citet{co87}, 
C92 = \citet{c92}, 
CH06 = \citet{ch06},
CL03 = \citet{cl03}, 
DEC03 = \citet{dec03},
FR96 = \citet{fr96},
HDJL01 = \citet{habing01}, 
HILL08 = \citet{hill08},
HO98 = \citet{ho98}, 
JJE86 = \citet{jje86}, 
JU04 = \citet{ju04}, 
MA05 = \citet{mamajek05}, 
MB98 = \citet{mannings-barlow98},
O92 = \citet{o92},
PAS06 = \citet{pas06}, 
PW91 = \citet{pw91},
RI05 = \citet{rieke05}, 
SB91 = \citet{sb91}, 
SCBS01 = \citet{scbs01},
SH03 = \citet{sh03},
SCH06 = \citet{sch06}, 
SN86 = \citet{sn86},
SU06 = \citet{su06}, 
TE00 = \citet{te00}, 
TR08 = \citet{trilling08}, 
WA95 = \citet{waters95}, 
WW88 = \citet{ww88},
WY07 = \citet{wyatt07a}.

\begin{longtable}{lrrrrr}
\caption{Metallicities, dispersion ($\delta$) and number of lines (N) used with the
WIDTH9 program, applying the N93 and C97 calibrations for the Vega-like sample}
\\
\hline \hline
      & N93 & N93 &  C97 & C97 & N\\
Star  & [Fe/H] & $\delta$[Fe/H] &  [Fe/H] & $\delta$[Fe/H] \\
\hline
\endfirsthead
\caption{Continued.}\\
\hline \hline
Star  & [Fe/H] & $\delta$[Fe/H] &  [Fe/H] & $\delta$[Fe/H] & N\\
\hline
\endhead
\hline
\endfoot
HD 105     &     $-$0.37 &     0.26 &     $-$0.33 &     0.26 &     15\\
HD 142     &     $-$0.45 &     0.27 &     $-$0.27 &     0.25 &     20\\
HD 2623    &     $-$0.20 &     0.26 &        0.09 &     0.20 &     23\\
HD 3003    &        0.17 &     0.22 &        0.07 &     0.31 &     17\\
HD 9672    &     $-$0.32 &     0.26 &     $-$0.31 &     0.21 &     24\\
HD 10647   &        0.12 &     0.22 &     $-$0.07 &     0.28 &     29\\
HD 10700   &     $-$0.73 &     0.29 &     $-$0.67 &     0.23 &     17\\
HD 10800   &        0.16 &     0.27 &        0.12 &     0.26 &     30\\
HD 17206   &     $-$0.22 &     0.27 &        0.03 &     0.21 &     22\\
HD 17848   &     $-$0.02 &     0.28 &     $-$0.17 &     0.20 &     21\\
HD 18978   &     $-$0.39 &     0.25 &     $-$0.11 &     0.23 &     22\\
HD 20010   &     $-$0.64 &     0.20 &     $-$0.62 &     0.30 &     27\\
HD 20794   &     $-$0.17 &     0.29 &     $-$0.58 &     0.24 &     17\\
HD 21563   &     $-$0.41 &     0.30 &     $-$0.10 &     0.26 &     19\\
HD 22049   &     $-$0.08 &     0.25 &     $-$0.13 &     0.27 &     22\\
HD 22484   &     $-$0.22 &     0.28 &     $-$0.19 &     0.26 &     21\\
HD 23362   &     $-$0.07 &     0.25 &     $-$0.47 &     0.29 &     25\\
HD 25457   &        0.18 &     0.21 &     $-$0.22 &     0.22 &     29\\
HD 28375   &        0.10 &     0.29 &     $-$0.19 &     0.26 &     23\\
HD 28978   &        0.20 &     0.22 &        0.33 &     0.24 &     18\\
HD 30495   &        0.11 &     0.29 &        0.13 &     0.24 &     23\\
HD 31295   &     $-$0.68 &     0.24 &     $-$0.76 &     0.25 &     27\\
HD 33262   &        0.07 &     0.25 &     $-$0.09 &     0.24 &     27\\
HD 33636   &        0.03 &     0.20 &     $-$0.09 &     0.27 &     16\\
HD 33949   &        0.00 &     0.21 &     $-$0.23 &     0.27 &     19\\
HD 35850   &     $-$0.23 &     0.27 &     $-$0.12 &     0.24 &     26\\
HD 36267   &     $-$0.23 &     0.23 &     $-$0.02 &     0.23 &     21\\
HD 37484   &     $-$0.17 &     0.31 &     $-$0.25 &     0.30 &     28\\
HD 38206   &     $-$0.06 &     0.21 &        0.32 &     0.23 &     23\\
HD 38385   &        0.09 &     0.25 &        0.12 &     0.30 &     24\\
HD 38393   &        0.30 &     0.21 &        0.21 &     0.27 &     26\\
HD 38678   &     $-$0.13 &     0.21 &     $-$0.35 &     0.28 &     29\\
HD 39014   &     $-$0.41 &     0.26 &     $-$0.39 &     0.30 &     23\\
HD 39060   &        0.00 &     0.29 &        0.17 &     0.21 &     16\\
HD 40136   &     $-$0.27 &     0.30 &     $-$0.33 &     0.27 &     29\\
HD 41700   &     $-$0.14 &     0.22 &     $-$0.41 &     0.22 &     23\\
HD 41742   &     $-$0.31 &     0.28 &     $-$0.30 &     0.30 &     16\\
HD 43955   &     $-$0.15 &     0.26 &     $-$0.20 &     0.26 &     21\\
HD 66591   &        0.04 &     0.25 &     $-$0.09 &     0.21 &     23\\
HD 68456   &     $-$0.36 &     0.26 &     $-$0.20 &     0.24 &     21\\
HD 69830   &     $-$0.07 &     0.23 &     $-$0.06 &     0.29 &     24\\
HD 71043   &        0.19 &     0.25 &     $-$0.14 &     0.25 &     23\\
HD 71155   &     $-$0.11 &     0.26 &        0.25 &     0.21 &     21\\
HD 75416   &        0.10 &     0.28 &        0.01 &     0.26 &     24\\
HD 76151   &     $-$0.07 &     0.26 &     $-$0.07 &     0.22 &     22\\
HD 79108   &     $-$0.10 &     0.28 &     $-$0.26 &     0.22 &     16\\
HD 80950   &     $-$0.29 &     0.31 &     $-$0.28 &     0.20 &     18\\
HD 82943   &        0.35 &     0.30 &        0.32 &     0.22 &     28\\
HD 86087   &        0.27 &     0.24 &     $-$0.07 &     0.28 &     29\\
HD 88955   &     $-$0.14 &     0.30 &        0.12 &     0.29 &     16\\
HD 98800   &     $-$0.05 &     0.25 &     $-$0.24 &     0.23 &     23\\
HD 99211   &     $-$0.15 &     0.25 &        0.14 &     0.23 &     26\\
HD 102647  &     $-$0.24 &     0.20 &     $-$0.07 &     0.27 &     27\\
HD 105211  &     $-$0.36 &     0.22 &     $-$0.04 &     0.26 &     18\\
HD 105686  &     $-$0.72 &     0.29 &     $-$0.39 &     0.27 &     13\\
HD 108257  &     $-$0.38 &     0.24 &     $-$0.36 &     0.21 &     16\\
HD 108483  &        0.02 &     0.21 &        0.14 &     0.25 &     20\\
HD 109085  &     $-$0.20 &     0.24 &     $-$0.23 &     0.27 &     20\\
HD 109573  &     $-$0.06 &     0.31 &        0.10 &     0.30 &     15\\
HD 111786  &     $-$1.42 &     0.30 &     $-$1.65 &     0.22 &     24\\
HD 113766  &        0.06 &     0.23 &     $-$0.14 &     0.27 &     16\\
HD 115617  &        0.20 &     0.29 &        0.13 &     0.24 &     16\\
HD 115892  &     $-$0.29 &     0.27 &     $-$0.33 &     0.28 &     14\\
HD 117176  &        0.01 &     0.26 &     $-$0.12 &     0.26 &     25\\
HD 117360  &     $-$0.39 &     0.21 &     $-$0.61 &     0.26 &     25\\
HD 121847  &     $-$0.30 &     0.26 &        0.10 &     0.27 &     16\\
HD 123160  &        0.27 &     0.27 &        0.17 &     0.25 &     28\\
HD 124771  &        0.15 &     0.21 &        0.01 &     0.21 &     29\\
HD 128311  &        0.16 &     0.24 &        0.04 &     0.29 &     16\\
HD 131885  &     $-$0.44 &     0.29 &     $-$0.19 &     0.28 &     23\\
HD 135344  &     $-$0.37 &     0.21 &     $-$0.41 &     0.28 &     20\\
HD 136246  &     $-$0.29 &     0.23 &     $-$0.49 &     0.28 &     27\\
HD 139365  &        0.02 &     0.25 &        0.35 &     0.29 &     26\\
HD 139664  &     $-$0.46 &     0.28 &     $-$0.08 &     0.26 &     24\\
HD 141569  &     $-$0.32 &     0.27 &     $-$0.01 &     0.20 &     18\\
HD 142096  &     $-$0.28 &     0.28 &     $-$0.16 &     0.23 &     14\\
HD 142114  &        0.09 &     0.26 &        0.06 &     0.26 &     21\\
HD 142165  &     $-$0.03 &     0.23 &        0.10 &     0.30 &     20\\
HD 144432  &     $-$0.19 &     0.23 &     $-$0.13 &     0.28 &     25\\
HD 145482  &     $-$0.38 &     0.27 &     $-$0.19 &     0.25 &     20\\
HD 150638  &     $-$0.34 &     0.23 &     $-$0.52 &     0.22 &     22\\
HD 152391  &     $-$0.24 &     0.26 &     $-$0.26 &     0.31 &     27\\
HD 158643  &     $-$0.44 &     0.22 &     $-$0.20 &     0.30 &     26\\
HD 158793  &        0.25 &     0.25 &        0.37 &     0.24 &     16\\
HD 159082  &        0.09 &     0.31 &     $-$0.22 &     0.24 &     26\\
HD 160691  &     $-$0.07 &     0.25 &     $-$0.06 &     0.21 &     21\\
HD 161868  &        0.11 &     0.30 &     $-$0.26 &     0.28 &     26\\
HD 164249  &     $-$0.04 &     0.22 &     $-$0.04 &     0.23 &     29\\
HD 164577  &     $-$0.14 &     0.24 &     $-$0.37 &     0.24 &     24\\
HD 165341  &     $-$0.26 &     0.30 &     $-$0.38 &     0.26 &     26\\
HD 166841  &     $-$0.16 &     0.31 &        0.05 &     0.27 &     16\\
HD 169830  &     $-$0.15 &     0.23 &        0.31 &     0.25 &     26\\
HD 176638  &     $-$0.21 &     0.23 &     $-$0.05 &     0.28 &     21\\
HD 177817  &        0.13 &     0.27 &        0.04 &     0.21 &     26\\
HD 178253  &        0.12 &     0.23 &     $-$0.26 &     0.22 &     23\\
HD 181296  &        0.06 &     0.30 &        0.14 &     0.28 &     21\\
HD 181327  &        0.34 &     0.23 &        0.24 &     0.28 &     27\\
HD 181869  &     $-$0.02 &     0.28 &     $-$0.03 &     0.29 &     19\\
HD 183324  &     $-$1.13 &     0.27 &     $-$1.29 &     0.30 &     21\\
HD 185507  &     $-$0.14 &     0.27 &     $-$0.07 &     0.21 &     31\\
HD 188228  &     $-$0.17 &     0.25 &     $-$0.02 &     0.31 &     17\\
HD 191089  &     $-$0.41 &     0.25 &     $-$0.17 &     0.29 &     24\\
HD 198160  &     $-$0.78 &     0.29 &     $-$0.99 &     0.24 &     25\\
HD 199260  &     $-$0.26 &     0.28 &     $-$0.01 &     0.31 &     18\\
HD 203608  &     $-$0.43 &     0.28 &     $-$0.72 &     0.24 &     19\\
HD 206893  &     $-$0.07 &     0.27 &        0.20 &     0.23 &     24\\
HD 207129  &     $-$0.26 &     0.20 &     $-$0.22 &     0.27 &     26\\
HD 209253  &        0.02 &     0.27 &     $-$0.26 &     0.26 &     16\\
HD 216435  &        0.07 &     0.26 &     $-$0.17 &     0.25 &     22\\
HD 216437  &        0.34 &     0.21 &        0.27 &     0.28 &     21\\
HD 216956  &     $-$0.42 &     0.28 &     $-$0.22 &     0.20 &     22\\
HD 221853  &     $-$0.18 &     0.24 &        0.08 &     0.26 &     15\\
HD 224392  &        0.09 &     0.21 &        0.14 &     0.27 &     23\\
\label{width.metal}
\end{longtable}

\begin{longtable}{lrcr}
\caption{T$_{\rm eff}$, Log g and [Fe/H] derived using the Downhill method, for the sample
of Vega-like stars}
\\
\hline \hline
Star  & T$_{\rm eff}$ [K] & Log g &  [Fe/H] \\
\hline
\endfirsthead
\caption{Continued.}\\
\hline \hline
Star  & T$_{\rm eff}$ [K] & Log g &  [Fe/H] \\
\hline
\endhead
\hline
\endfoot
HD 105     &      5989 &     4.57 &     $-$0.13\\
HD 142     &      6182 &     4.12 &     $-$0.21\\
HD 2623    &      4923 &     4.10 &     $-$0.15\\
HD 3003    &      8794 &     4.10 &        0.06\\
HD 9672    &      8865 &     4.21 &     $-$0.12\\
HD 10647   &      5954 &     4.67 &     $-$0.01\\
HD 10700   &      5499 &     4.97 &     $-$0.53\\
HD 10800   &      5901 &     4.84 &        0.09\\
HD 17206   &      6359 &     4.57 &     $-$0.14\\
HD 17848   &      8308 &     3.96 &     $-$0.05\\
HD 18978   &      8050 &     4.14 &     $-$0.25\\
HD 20010   &      6072 &     4.07 &     $-$0.39\\
HD 20794   &      5629 &     4.70 &     $-$0.35\\
HD 21563   &      6714 &     4.22 &     $-$0.25\\
HD 22049   &      4963 &     3.92 &     $-$0.17\\
HD 22484   &      5943 &     4.29 &     $-$0.17\\
HD 23362   &      4899 &     4.21 &     $-$0.23\\
HD 25457   &      6364 &     4.68 &        0.00\\
HD 28375   &     15275 &     4.20 &     $-$0.02\\
HD 28978   &      9075 &     4.26 &        0.17\\
HD 30495   &      5759 &     4.53 &        0.01\\
HD 31295   &      8651 &     4.11 &     $-$0.75\\
HD 33262   &      6073 &     4.83 &     $-$0.10\\
HD 33636   &      5744 &     4.56 &     $-$0.08\\
HD 33949   &     12459 &     3.44 &     $-$0.07\\
HD 35850   &      6021 &     4.66 &     $-$0.05\\
HD 36267   &     14760 &     4.27 &     $-$0.01\\
HD 37484   &      6380 &     4.54 &     $-$0.22\\
HD 38206   &     10135 &     4.36 &        0.14\\
HD 38385   &      6726 &     3.87 &        0.02\\
HD 38393   &      6163 &     4.37 &        0.08\\
HD 38678   &      8327 &     3.97 &     $-$0.19\\
HD 39014   &      7489 &     3.41 &     $-$0.40\\
HD 39060   &      8036 &     4.21 &        0.11\\
HD 40136   &      7007 &     4.12 &     $-$0.32\\
HD 41700   &      6079 &     4.55 &     $-$0.22\\
HD 41742   &      6331 &     4.61 &     $-$0.33\\
HD 43955   &     17890 &     4.12 &     $-$0.15\\
HD 66591   &     16641 &     4.15 &     $-$0.19\\
HD 68456   &      6305 &     4.14 &     $-$0.39\\
HD 69830   &      5586 &     5.15 &        0.16\\
HD 71043   &     10103 &     4.31 &     $-$0.02\\
HD 71155   &      9881 &     4.22 &        0.14\\
HD 75416   &     12603 &     4.25 &        0.16\\
HD 76151   &      5750 &     4.46 &     $-$0.16\\
HD 79108   &     10273 &     4.11 &     $-$0.07\\
HD 80950   &     10330 &     4.36 &     $-$0.05\\
HD 82943   &      5764 &     4.25 &        0.30\\
HD 86087   &      9310 &     4.25 &        0.08\\
HD 88955   &      8707 &     4.04 &     $-$0.02\\
HD 98800   &      4595 &     3.99 &     $-$0.22\\
HD 99211   &     10625 &     4.90 &     $-$0.01\\
HD 102647  &      8522 &     4.26 &     $-$0.25\\
HD 105211  &      6901 &     3.91 &     $-$0.29\\
HD 105686  &      9930 &     4.19 &     $-$0.48\\
HD 108257  &     16576 &     3.98 &     $-$0.53\\
HD 108483  &     20320 &     4.33 &     $-$0.06\\
HD 109085  &      6756 &     4.17 &     $-$0.21\\
HD 109573  &      9378 &     4.43 &     $-$0.03\\
HD 111786  &      8115 &     3.84 &     $-$1.45\\
HD 113766  &      6796 &     4.32 &        0.09\\
HD 115617  &      5558 &     4.55 &        0.07\\
HD 115892  &      8600 &     4.11 &     $-$0.46\\
HD 117176  &      5495 &     4.02 &     $-$0.08\\
HD 117360  &      6314 &     4.51 &     $-$0.45\\
HD 121847  &     12472 &     4.00 &     $-$0.09\\
HD 123160  &      4356 &     4.10 &        0.04\\
HD 124771  &     16136 &     4.18 &     $-$0.02\\
HD 128311  &      4635 &     4.71 &     $-$0.04\\
HD 131885  &      9680 &     4.20 &     $-$0.23\\
HD 135344  &      6692 &     4.11 &     $-$0.20\\
HD 136246  &      9790 &     4.30 &     $-$0.28\\
HD 139365  &     17990 &     4.33 &        0.17\\
HD 139664  &      6693 &     4.55 &     $-$0.31\\
HD 141569  &      9963 &     4.11 &     $-$0.07\\
HD 142096  &     17034 &     4.75 &     $-$0.27\\
HD 142114  &     18429 &     4.42 &        0.23\\
HD 142165  &     14077 &     4.31 &        0.11\\
HD 144432  &      6957 &     3.55 &     $-$0.18\\
HD 145482  &     19214 &     4.32 &     $-$0.24\\
HD 150638  &     12453 &     4.16 &     $-$0.42\\
HD 152391  &      5418 &     5.05 &     $-$0.12\\
HD 158643  &      9772 &     3.12 &     $-$0.25\\
HD 158793  &      9781 &     3.03 &        0.32\\
HD 159082  &     10990 &     3.91 &     $-$0.06\\
HD 160691  &      5600 &     4.30 &        0.09\\
HD 161868  &      8567 &     3.98 &     $-$0.06\\
HD 164249  &      6620 &     4.32 &     $-$0.09\\
HD 164577  &      9687 &     3.67 &     $-$0.29\\
HD 165341  &      5153 &     4.20 &     $-$0.32\\
HD 166841  &     10885 &     3.36 &     $-$0.02\\
HD 169830  &      6349 &     4.08 &        0.08\\
HD 176638  &     10095 &     4.10 &     $-$0.21\\
HD 177817  &     12667 &     3.72 &        0.19\\
HD 178253  &      8448 &     4.01 &     $-$0.11\\
HD 181296  &      9207 &     4.30 &        0.17\\
HD 181327  &      6449 &     4.44 &        0.29\\
HD 181869  &     12100 &     4.00 &        0.18\\
HD 183324  &     10325 &     4.17 &     $-$1.24\\
HD 185507  &     21374 &     4.59 &        0.04\\
HD 188228  &     10366 &     4.23 &     $-$0.13\\
HD 191089  &      6402 &     4.33 &     $-$0.34\\
HD 198160  &      7860 &     4.02 &     $-$1.03\\
HD 199260  &      6231 &     4.37 &     $-$0.11\\
HD 203608  &      6105 &     4.61 &     $-$0.51\\
HD 206893  &      6454 &     4.40 &     $-$0.05\\
HD 207129  &      5776 &     4.39 &     $-$0.12\\
HD 209253  &      6175 &     4.62 &     $-$0.17\\
HD 216435  &      5755 &     3.82 &     $-$0.17\\
HD 216437  &      5757 &     3.99 &        0.20\\
HD 216956  &      8743 &     4.09 &     $-$0.34\\
HD 221853  &      6196 &     4.02 &        0.00\\
HD 224392  &      8778 &     4.06 &        0.07\\
\label{downhill.metal}
\end{longtable}

\end{document}